\documentclass[prb,fleqn,12pt,showpacs]{revtex4}
\usepackage{epsfig}
\usepackage{graphicx}
\usepackage{amsfonts}
\usepackage{amsmath}
\begin{document}
\title{Onset and melting of local orbital order}
\author{Avinash Singh and Dheeraj Kumar Singh}
\email{avinas@iitk.ac.in}
\affiliation{Department of Physics, Indian Institute of Technology Kanpur - 208016}
\begin{abstract}
The onset and melting of locally staggered charge/orbital correlations is investigated within a two-orbital correlated electron model with inter-orbital and inter-site Coulomb interactions. Beyond the critical doping concentration $x_c \approx 0.2$, the $(\pi,\pi,\pi)$ staggered orbital order in the ferromagnet re-emerges sharply at finite temperature. The CE-type orbital correlation exhibits a sharp onset close to the Curie temperature and rapid thermal melting thereafter, which provides quantitative understanding of the $(\pi/2,\pi/2,0)$ feature observed in neutron scattering experiments on 
$\rm La_{0.7} (Ca_{y} Sr_{1-y})_{0.3} Mn O_3$ single crystals. In the zig-zag AF state, the CE-type orbital correlations are found to be even more readily stabilized, but only within a narrow doping regime around $x=0.5$. 
\end{abstract}
\pacs{75.30.Ds,71.27.+a,75.10.Lp,71.10.Fd}
\maketitle
\newpage
\section{Introduction}
The orbital degree of freedom of the electron has attracted considerable attention in recent years due to the rich variety of electronic, magnetic, and transport properties exhibited by orbitally degenerate systems such as the ferromagnetic manganites, which have highlighted the interplay between spin and orbital degrees of freedom in these correlated electron systems.\cite{tokura_2003,khaliullin_2005} Orbital fluctuations, correlations, and orderings have been observed in Raman spectroscopic studies\cite{saitoh_2001} of orbiton modes in $\rm LaMnO_3$, polarization-contrast-microscopy studies\cite{ogasawara_2001} of $\rm La_{0.5}Sr_{1.5}MnO_4$, magnetic susceptibility and inelastic neutron scattering studies\cite{khalifah_2002} of $\rm La_4 Ru_2 O_{10}$, and resonant inelastic soft X-ray scattering studies\cite{ulrich_2008} of $\rm YTiO_3$ and $\rm LaTiO_3$. A new detection method for orbital structures and ordering based on spectroscopic imaging scanning tunneling microscopy is of strong current interest\cite{lee_2009} in orbitally active metallic systems such as strontium ruthenates and iron pnictide superconductors.

A composite charge-orbital ordering is exhibited by half-doped narrow-bandwidth manganites such as $\rm La_{0.5} Ca_{0.5} Mn O_3$, with nominally Mn$^{3+}$ and Mn$^{4+}$ atoms on alternate lattice sites in a checkerboard pattern, a staggered orbital ordering of the Mn$^{3+}$ electron between the two $e_g$ orbitals corresponding to wavevector $(\pi/2,\pi/2,0)$, and a CE-type AF ordering of the Mn core spins arranged in ferromagnetic zig-zag chains.\cite{wollan_1955,goodenough_1955} 

Such CE-type orbital correlations persist in the metallic ferromagnetic phase of the colossal magnetoresistive (CMR) manganites, as revealed in neutron and x-ray scattering experiments in ferromagnetic bilayer \cite{argyriou_2002,campbell_2003} and pseudo-cubic manganites.\cite{adams_2000,lynn_2007} Short-range (10-20\AA) charge and orbital correlations associated with the CMR in the orthorhombic paramagnetic phase were observed as diffuse peaks at wave vector $(\pi/2,\pi/2,0)$ in the same positions as superlattice peaks in the CE-type charge-orbital structure. 

Sharp onset of CE-type dynamical charge/orbital correlations near the Curie temperature were also indicated in Raman scattering studies of $\rm Sm_{1-x} Sr_x Mn O_3$ $(x=0.45)$ single crystals.\cite{saitoh_2002} Systematic enhancement of this feature was observed with bandwidth reduction, along with the enhancement of CMR. CE-type dynamical charge/orbital correlations and accompanying collective and dynamical lattice distortions were suggested as being responsible for the steep metal-insulator transition and CMR near the charge/orbital ordering instability. 

Recent neutron scattering studies of $\rm La_{1-x}(Ca_y Sr_{1-y})_x MnO_3$ crystals have also revealed a sharp onset of the $(\pi/2,\pi/2,0)$ diffuse peak near the Curie temperature,\cite{moussa_2007} which gradually diminishes in intensity and sharpness with decreasing Ca concentration. This behaviour of CE-type orbital correlations is remarkably similar to the resistivity temperature profile of these crystals with varying Ca concentration,\cite{tomioka_2000} and the sharp rise in resistivity and the metal-insulator transition observed near the Curie temperature.

CE-type orbital correlations are also important for magnetic couplings and excitations in the ferromagnetic manganites. Due to a spin-orbital coupling effect, low-energy staggered orbital fluctuation modes, particularly with momentum near $(\pi/2,\pi/2,0)$, generically yield strong intrinsically non-Heisenberg $(1-\cos q)^2$ magnon self energy correction, resulting in no spin stiffness reduction, but strongly suppressed zone-boundary magnon energies in the $\Gamma$-X direction,\cite{sporb,qfklm2} and can quantitatively account for the several zone-boundary anomalies observed in spin-wave excitation measurements on ferromagnetic manganites.\cite{hwang_98,dai_2000,tapan_2002,ye_2006,ye_2007,zhang_2007,moussa_2007} 

Although the Jahn-Teller electron-phonon coupling is considered to be important in manganites, especially in the low and intermediate doping range,\cite{millis_1995} the inter-orbital Coulomb interaction has been suggested to be much stronger than the electron-phonon coupling in order to account for the observed insulating behaviour in undoped manganites above the Jahn-Teller transition and the bond length changes below it.\cite{benedetti_1999,okamoto_2002} Generally, Coulombic and Jahn-Teller-phononic approaches for manganites have been shown to be qualitatively similar.\cite{hotta_2000} Indeed, a mean-field treatment of the Jahn-Teller term\cite{stier_2007} yields an electronic exchange field in orbital space proportional to the orbital magnetization $\langle n_{i\sigma\alpha} -  n_{i\sigma\beta}\rangle$, exactly as would be obtained from the inter-orbital interaction term. Since orbital correlations are driven by both inter-orbital Coulomb interaction as well as Jahn-Teller electron-phonon coupling, local orbital correlations are therefore generally accompanied with local lattice distortion.  

While the role of inter-orbital Coulomb interaction has been investigated for the undoped ($x=0$) parent compound $\rm LaMnO_3$,\cite{benedetti_1999,okamoto_2002} and at half doping ($x=0.5$),\cite{hotta_2000} surprisingly the thermal onset and melting of different types of local orbital correlations in the full doping range $0 < x < 0.5$ for ferromagnetic manganites has not been studied within correlated electron models. This should be of much interest in view of the importance of orbital correlations on spin dynamics, their observation close to $T_c$ in neutron, X-ray, and Raman scattering experiments, and their likely role in the observed finite-temperature metal-insulator transition and the CMR effect. 

In this paper, we will therefore investigate the doping dependence of local orbital correlations and their sensitivity on the inter-orbital ($V$) and inter-site $(V')$ Coulomb interactions within a two-orbital interacting electron model. Due to band narrowing in the correlated ferromagnet, a strong sensitivity on the ratio $V/t$ would result in sharp onset of local orbital correlations near the Curie temperature. A detailed comparison of the overall shape with observed neutron scattering features can provide fundamental insight into onset and melting of local orbital correlations in manganites. We will consider staggered $(\pi,\pi,\pi)$ orbital correlations, CE-type orbital correlations in the ferromagnetic state, and CE-type orbital correlations in the zig-zag AF state, which are of interest in different doping regimes of manganites.

We will be concerned here only with local orbital correlations and not long-range orbital order. We will therefore confine our investigation to the Hartree-Fock (HF) level which correctly describes the local orbital moment formation associated with the short time scale correlations corresponding to the high interaction energy scale $V$. Earlier HF studies of two-orbital interacting electron models have obtained qualitatively correct phase diagram for electron-doped ($x \ge 0.5$) manganites.\cite{maitra_2003}

\section{Staggered orbital order}

The undoped parent compound $\rm LaMnO_3$ has a staggered ($\pi,\pi,0$) orbital structure with antiferro-orbital ordering in the ferromagnetic plane and ferro-orbital ordering in the perpendicular direction. Weakly doped manganites, which are ferromagnetic insulators, also exhibit an orbitally ordered state for $x \lesssim 0.2$, as inferred from x-ray diffraction and neutron scattering experiments.\cite{pissas_2005,moussa_2007} 

In this section, we will examine the temperature dependence of the local staggered orbital ordering in the low-doping regime. We therefore consider the two orbital model: 
\begin{equation}
H = -t \sum_{\langle ij \rangle \sigma\mu} a^\dagger _{i\sigma\mu}  a_{j\sigma\mu} - J \sum_{i\mu} {\bf S}_{i\mu}.{\mbox{\boldmath $\sigma$}}_{i\mu} + V \sum_i n_{i\alpha} n_{i\beta} 
\end{equation}
corresponding to the two $e_g$ orbitals $\mu=\alpha,\beta$ per site. An inter-orbital ($V$) Coulomb interaction is included along with the local exchange interaction $J$ between the localized Mn core spins ${\bf S}_i$ and itinerant electron spins ${\mbox{\boldmath $\sigma$}}_i$. 

For simplicity, we consider an orbitally-ordered ferromagnetic state having staggered orbital ordering in all three directions, with (spin-$\uparrow$) electronic densities:
\begin{eqnarray}
\langle n_{\alpha}\rangle_A &=& \langle n_{\beta}\rangle_B = n + \delta m/2 \nonumber \\
\langle n_{\beta}\rangle_A &=& \langle n_{\alpha}\rangle_B = n - \delta m/2 
\end{eqnarray}
for the two orbitals $\alpha$ and $\beta$, where $\delta m$ is the staggered orbital order and characterizes the density modulation on the two sublattices A and B. The spin-$\downarrow$ electronic densities vanish as the spin-$\downarrow$ electronic bands are shifted above the Fermi energy by the exchange splitting $2JS$ in the ferromagnetic state. In the pseudo-spin space of the two orbitals, this orbital density wave (ODW) state is exactly analogous to the antiferromagnetic state of the Hubbard model with staggered spin ordering.\cite{as_af_ordering} The self-consistent field approximation results in identical effective single-particle Hamiltonian in the two-sublattice basis: 
\begin{equation}
H^{(0)} _\mu ({\bf k}) =  \begin{pmatrix} -\mu\Delta  & \epsilon_{\bf k} \\ 
\epsilon_{\bf k} & \mu\Delta  \end{pmatrix}
\end{equation}
in terms of the self-consistently determined orbital exchange field $\Delta = V\delta m/2$, with $\mu = \pm$ for the two orbitals $\alpha/\beta$. 

A similar expression is obtained in the mean-field approximation of the Jahn-Teller part involving the coupling of $e_g$ electrons with JT phonon modes,\cite{hotta_2000,stier_2007} with the equivalence $g^2/k = V/2$. For a value $\lambda \equiv g/\sqrt{kt} \sim 1.4$ of the dimensionless coupling constant, which lies in the range considered for manganites, the corresponding effective interaction energy $V/t = 2g^2/kt = 2\lambda ^2 \sim 4$, as considered in our recent detailed comparison of calculated spin dynamics for manganites with experiments.\cite{qfklm2} The eigenvalues and eigenvectors of the above Hamiltonian matrix yield the bare-level band-electron energies and amplitudes:
\begin{eqnarray}
E_{{\bf k}\mu} &=& \pm \sqrt{\Delta^2 + \epsilon_{\bf k}^2} \nonumber \\
a_{{\bf k}\alpha\ominus}^2 &=& \frac{1}{2}\left (1 + \frac{\Delta}{\sqrt{\Delta^2 + \epsilon_{\bf k}^2}}\right ) = b_{{\bf k}\alpha\oplus}^2 = b_{{\bf k}\beta\ominus}^2 \nonumber \\
b_{{\bf k}\alpha\ominus}^2 &=& \frac{1}{2}\left (1 - \frac{\Delta}{\sqrt{\Delta^2 + \epsilon_{\bf k}^2}}\right ) = a_{{\bf k}\alpha\oplus}^2 = a_{{\bf k}\beta\ominus}^2
\end{eqnarray}
for orbitals $\mu = \alpha, \beta$, where $\oplus$ and $\ominus$ refer to the two eigenvalue branches ($\pm$). Orbital ordering splits the electron bands with energy gap $2\Delta = V\delta m$. 

An important characteristic of the ODW state, as seen from equation (4), is that states near the top of the band $(\epsilon_{\bf k} = 0$)  contribute dominantly to orbital ordering (strongly "orbitic"), with densities 1 and 0 on the two sublattices for orbital $\mu=\alpha$, whereas states deep in the band $(|\epsilon_{\bf k}| \gg \Delta$) are nearly non-orbitic, with densities 1/2 on both sublattices. 

Similarity with the AF state of the Hubbard model also results in identical self-consistency condition:
\begin{equation}
\delta m = \sum_{\bf k} \frac{\Delta}{\sqrt{\Delta^2 + \epsilon_{\bf k}^2}}(f^- - f^+)
\end{equation}
where $f^-(T)$ and $f^+(T)$ are the temperature-dependent Fermi functions corresponding to the two branches, and the chemical potential is determined from the total fermion density constraint: 
\begin{equation}
n=1-x=\sum_{\bf k} (f^- + f^+) \; .
\end{equation}
Coupled equations (5) and (6) then self consistently determine the magnitude of the local orbital order $\delta m=2\Delta/V$ as a function of the effective interaction strength $V$ and temperature $T$. When only nearest-neighbor hopping is present, orbital ordering at quarter filling sets in for any positive $V$ due to Fermi-surface nesting, whereas a finite critical interaction strength is required when frustrating next-nearest-neighbor hopping terms are included.

\begin{figure}
\begin{center}
\vspace*{-2mm}
\hspace*{-0mm}
\psfig{figure=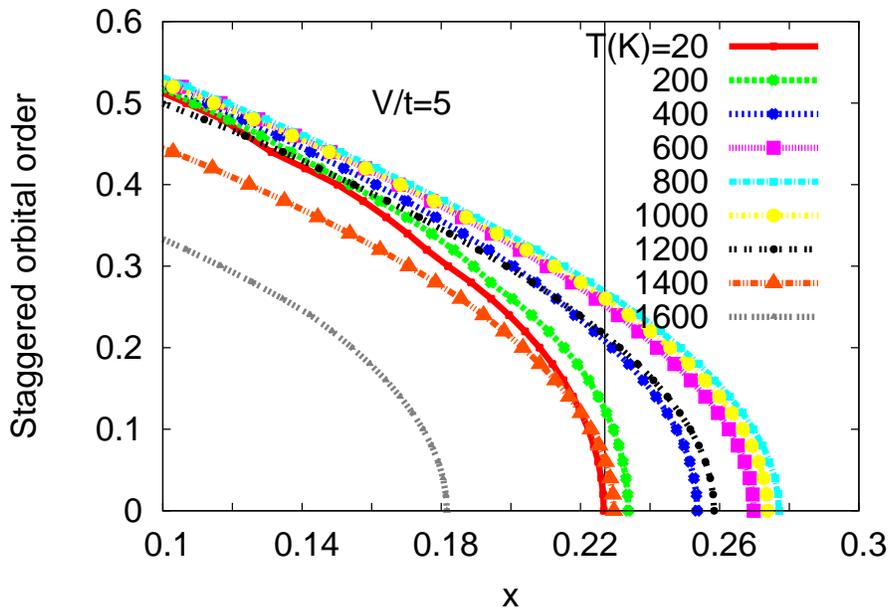,width=120mm}
\vspace{-5mm}
\end{center}
\caption{Reduction of local staggered $(\pi,\pi,\pi)$ orbital order with doping, shown at different temperatures. Beyond the zero-temperature critical doping concentration $x_c \approx 0.22$ (shown by a vertical line), orbital order reappears sharply at finite temperature.}
\label{mvsx}
\end{figure}

\begin{figure}
\begin{center}
\vspace*{-2mm}
\hspace*{-0mm}
\psfig{figure=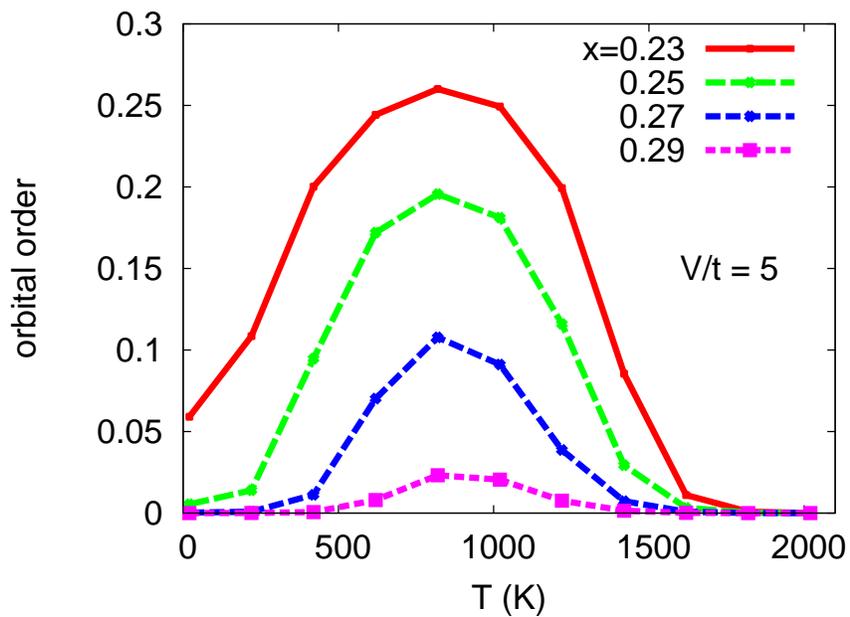,width=120mm}
\vspace{-5mm}
\end{center}
\caption{Temperature dependence of the local staggered $(\pi,\pi,\pi)$ orbital order at fixed doping $x$ and interaction strength, showing sharp onset and melting with temperature.}
\label{mvst}
\end{figure}

Fig. \ref{mvsx} shows the reduction of local staggered $(\pi,\pi,\pi)$ orbital order with doping $(x)$ at different temperatures. Here the hopping energy scale $t=200$meV $\approx 2000$K. At low temperature, as the strongly orbitic electronic states at the top of the lower band are progressively emptied upon doping, local orbital order decreases rapidly and eventually vanishes at a critical doping concentration $x_c \approx 0.22$, shown by a vertical line. Beyond this critical doping, local orbital order re-emerges sharply at finite temperature due to thermal occupation of these strongly orbitic states, as shown in Fig. \ref{mvst}, and then gradually melts away at high temperature. This onset of orbital order will be further enhanced if the spin-disordering induced bandwidth reduction near the Curie temperature is included, as discussed in the next section.

The sharp onset of staggered $(\pi)$ orbital correlations below $x_c \sim 0.2$ results in sharply reduced magnon energies and Curie temperature, as obtained in investigation of spin dynamics in the orbitally ordered state, which is in good agreement with experiments.\cite{qfklm2} Also, the measured spin stiffness\cite{moussa_2007} in the low bandwidth (high $V/t$) compound $\rm La_{1-x} Ca_{x} Mn O_3$ shows a jump at $x_c =0.22$ separating the ferromagnetic insulating and metallic phases, whereas the high bandwidth (low $V/t$) compound $\rm La_{1-x} Sr_{x} Mn O_3$ shows a much weaker discontinuity at much lower $x_c =0.17$.  

Although long-range orbital order will be susceptible to orbital fluctuations at finite doping in the same way as for the staggered spin ordering in the doped antiferromagnet,\cite{as_doped_af} this HF result does provide a measure of the local orbital order. Fig. \ref{mvsx} also yields the doping dependence of the energy gap $2\Delta = m V$ corresponding to the local staggered orbital order. The gap decreases rapidly with doping and vanishes at critical doping concentration $x_c$, is roughly 1000K for $m=0.1$, $V/t=5$, $t=200$meV, and will be reduced significantly on including the electron-orbiton coupling and multiple orbiton emission/absorption processes. Beyond $x_c$, this gap reappears at finite temperature, leading to a metal-insulator transition and enhanced resistivity. This behaviour of the gap is similar to the observed pseudogap behaviour in cuprates and manganites.  


\section{CE-type orbital order in the ferromagnetic state}

At half doping, narrow bandwidth manganites such as $\rm La_{0.5}Ca_{0.5}MnO_3$ exhibit a CE-type antiferromagnetic order consisting of ferromagnetic zigzag chains with staggered charge and orbital ordering. Local CE-type orbital correlations also emerge in the ferromagnetic phase near the Curie temperature, coincident with the metal-insulator transition and the CMR effect, as mentioned earlier. While CE-type charge and orbital ordered ferromagnetic phase was studied at half doping using cooperative phonons with large electron-phonon coupling,\cite{hotta_2001,cepas_2006,schlottmann_2009,sboychakov_2010} and recently at hole density $x=1/4$ using Monte Carlo investigations,\cite{shen_2010} and at several fractional hole densities using numerical simulations on finite 3d clusters,\cite{rosciszewski_2010} the doping and temperature dependence of CE-type orbital correlations within an interacting electron model has not been investigated in the ferromagnetic phase. 

In order to investigate the onset and melting of combined charge-orbital orderings in the ferromagnetic state, we therefore consider the following two-orbital model corresponding to the two $e_g$ orbitals $\mu=\alpha,\beta$ per site:
\begin{equation}
H = -t \sum_{\langle ij \rangle \sigma\mu}a^\dagger_{i\sigma\mu}  a_{j\sigma\mu}-J \sum_{i\mu} {\bf S}_{i\mu}.{\mbox{\boldmath $\sigma$}}_{i\mu} + V \sum_i n_{i\alpha} n_{i\beta} 
+ V^{\prime} \sum_{\langle ij \rangle} n_{i} n_{j}
\end{equation}
including inter-orbital ($V$) and inter-site ($V'$) Coulomb interactions which will induce local orbital and charge correlations. For simplicity, we consider a staggered charge-orbital ordering as shown in Fig. \ref{ch_orb}, corresponding to ordering wavevector $\pi$ for charge order and $\pi/2$ for orbital order in all directions, with (spin-$\uparrow$) electronic densities:
\begin{eqnarray}
\langle n_{\alpha}\rangle_A &=& \langle n_{\beta}\rangle_C = (n + \delta n /2 +\delta m)/2 \nonumber \\
\langle n_{\alpha}\rangle_C &=& \langle n_{\beta}\rangle_A = (n + \delta n /2 -\delta m)/2 \nonumber\\
\langle n_{\alpha}\rangle_B &=& \langle n_{\alpha}\rangle_D = (n - \delta n /2)/2 \nonumber \\
\langle n_{\beta}\rangle_B &=& \langle n_{\beta} \rangle_D  = (n - \delta n /2)/2 
\end{eqnarray}
where $\delta n $ and $\delta m$ represent charge and orbital density modulation. The spin-$\downarrow$ electronic densities vanish at $T=0$ as the spin-$\downarrow$ bands are shifted above the Fermi energy by the exchange splitting $2JS$. In a four-sublattice basis ($\nu=A, B, C, D$), with the electron field operator $\Psi_{{\bf k}\mu} \equiv (a_{{\bf k}\mu} ^A \; a_{{\bf k}\mu} ^B \; a_{{\bf k}\mu} ^C\; a_{{\bf k}\mu} ^D)$, the HF-level Hamiltonian for spin-$\uparrow$ electrons:

\begin{figure}
\begin{center}
\vspace*{-2mm}
\hspace*{-0mm}
\psfig{figure=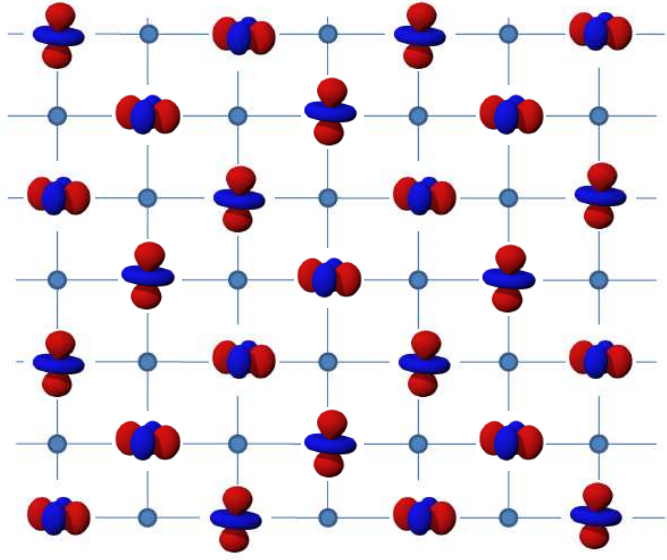,width=120mm}
\end{center}
\vspace{-25mm}
\caption{CE-type staggered charge/orbital correlations corresponding to ordering wavevectors $\pi$ and $\pi/2$ for charge and orbital orders, respectively.}
\label{ch_orb}
\end{figure}

\begin{equation}
H^{(0)} = \sum_{{\bf k},\mu} \Psi_{{\bf k}\mu} ^\dagger 
\begin{pmatrix} 
\Delta_{\rm ch} - \mu \Delta_{\rm orb} & \delta_{\bf k} & 0 & \delta^{\star}_{\bf k} \\
\delta^{\star}_{\bf k} & -\Delta_{\rm ch} & \delta_{\bf k} & 0 \\
0 & \delta^{\star}_{\bf k} & \Delta_{\rm ch} + \mu \Delta_{\rm orb}  & \delta_{\bf k} \\
\delta_{\bf k} & 0 & \delta^{\star}_{\bf k} & -\Delta_{\rm ch} 
\end{pmatrix} 
\Psi_{{\bf k}\mu}
\end{equation}
where the effective charge and orbital exchange fields ($z$ is the lattice coordination number):
\begin{eqnarray}
\Delta_{\rm ch} & = & (V/2 - zV')\delta n /2 \nonumber \\
\Delta_{\rm orb} & = & V \delta m/2 
\end{eqnarray}
and the NN hopping term (which mixes $AB,BC,CD$ and $DA$ sublattices): 
\begin{equation}
\delta_{\bf k} = -t (e^{ik_x} + e^{ik_y}+e^{ik_z}) \; .
\end{equation}

The Hamiltonian matrix (9) was numerically diagonalized to obtain the four eigenvalues $E_{{\bf k}\lambda}$ corresponding to the four sub-bands $\lambda$. The four-component eigenvectors $\phi_{{\bf k}\lambda} ^\nu$ yield the electronic amplitude on sublattice $\nu$. Evaluation of the new staggered charge and orbital order in terms of the electronic densities obtained by summing over occupied states yields the self-consistency conditions:
\begin{eqnarray}
\delta m &=& \sum^{E_{{\bf k}\lambda} < E_{\rm F}}_{{\bf k},\lambda}
|\phi_{{\bf k}\lambda} ^A|^2 - |\phi_{{\bf k}\lambda} ^C|^2 \nonumber \\ 
\delta n &=& \sum^{E_{{\bf k}\lambda} < E_{\rm F}}_{{\bf k},\lambda}
|\phi_{{\bf k}\lambda} ^A|^2 + |\phi_{{\bf k}\lambda} ^C|^2 - |\phi_{{\bf k}\lambda} ^B|^2 - |\phi_{{\bf k}\lambda} ^D|^2.
\end{eqnarray}
where the Fermi energy $E_{\rm F}$ is also determined self-consistently in terms of the average (spin-$\uparrow$) electronic density $n = 1-x$. The above self-consistent procedure provides an unrestricted HF scheme, with independent determination of charge and orbital order.  

\begin{figure}
\begin{center}
\vspace*{-2mm}
\hspace*{-0mm}
\psfig{figure=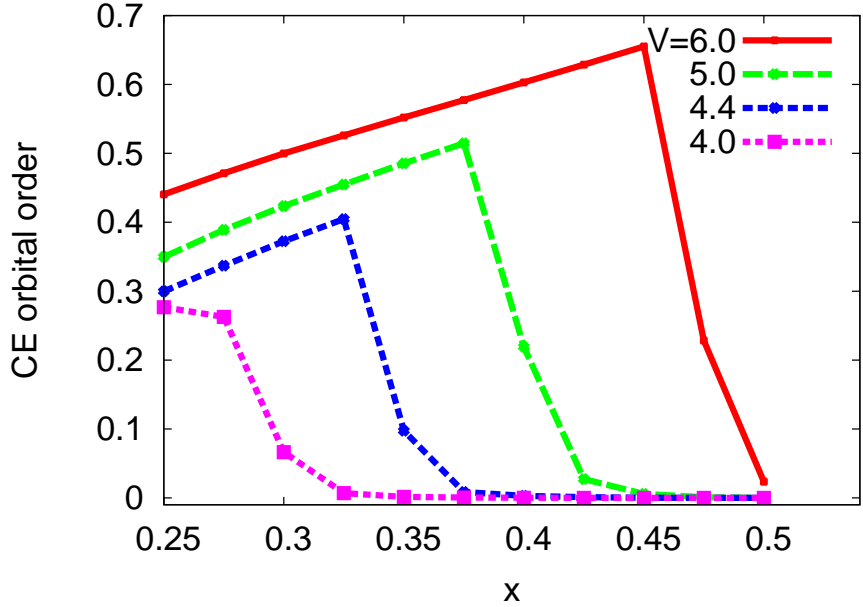,width=120mm}
\vspace{-5mm}
\end{center}
\caption{Behaviour of the CE-type orbital order with hole doping $x$ for different inter-orbital interaction strength $V$,
showing its strong sensitivity to small changes in $V/t$. Here $V'/t=0.73$.}
\label{fig2}
\end{figure}

\begin{figure}
\begin{center}
\vspace*{-2mm}
\hspace*{-0mm}
\psfig{figure=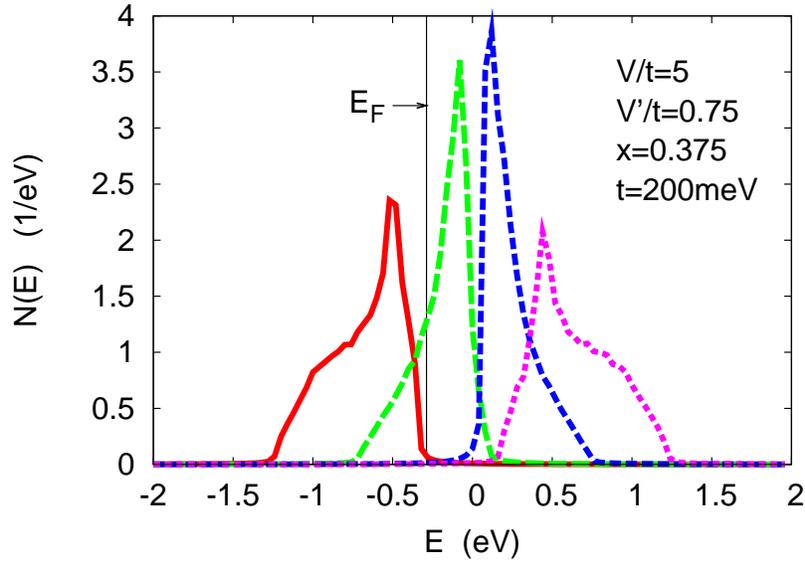,width=110mm}
\vspace{-5mm}
\end{center}
\caption{The density of states showing the four sub-bands in the self-consistent CE-type orbitally ordered ferromagnetic state.}
\label{dos_ferro}
\end{figure}

Fig. \ref{fig2} shows the behaviour of the self-consistent CE-type orbital order in the ferromagnetic state with hole doping $x$ for different inter-orbital interaction strength $V$. The orbital order is optimal around $x=0.4$ and decreases sharply towards half doping ($x=0.5$). Due to the small overlap between the two lowest-energy sub-bands as shown in Fig. \ref{dos_ferro}, the lowest sub-band is fully occupied not at $x=0.5$ but at slightly lower $x$. As the Fermi energy moves down with increasing $x$ towards 0.5, the maximally orbitic states at the top of the lowest sub-band get emptied, leading to a sharp suppression of the CE orbital order. For $0.3 < x < 0.35$, the orbital order rises sharply for a small change in $V/t$ from 4 to 4.4, rendering the system extremely sensitive to small changes in bandwidth and temperature. 

The strong presence of CE-type orbital correlations found here in the ferromagnetic state near $x=0.45$ is consistent with the observation of anomalous zone-boundary magnon softening\cite{zhang_2007}, which has been ascribed to correlation-induced magnon self energy correction arising from coupling of spin fluctuations with specifically CE-type orbital 
correlations.\cite{sporb,qfklm2}

A strong electron-hole asymmetry around half doping ($x=0.5$) directly follows from the strongly asymmetric band structure and orbitic character about the Fermi energy. While the CE-type orbital order is strongly enhanced for $x < 0.5$ due to added electrons going in the maximally orbitic states at the top of the lowest sub-band, it is rapidly suppressed due to further depletion of these states for $x > 0.5$. The CE-type orbital order peaks at $x \lesssim 0.5$ when the lowest sub-band gets completely filled, and further electron filling of opposite sublattice state at the bottom of second sub-band results in slow decrease of orbital order with decreasing $x$. 

\begin{figure}
\begin{center}
\vspace*{-2mm}
\hspace*{-0mm}
\psfig{figure=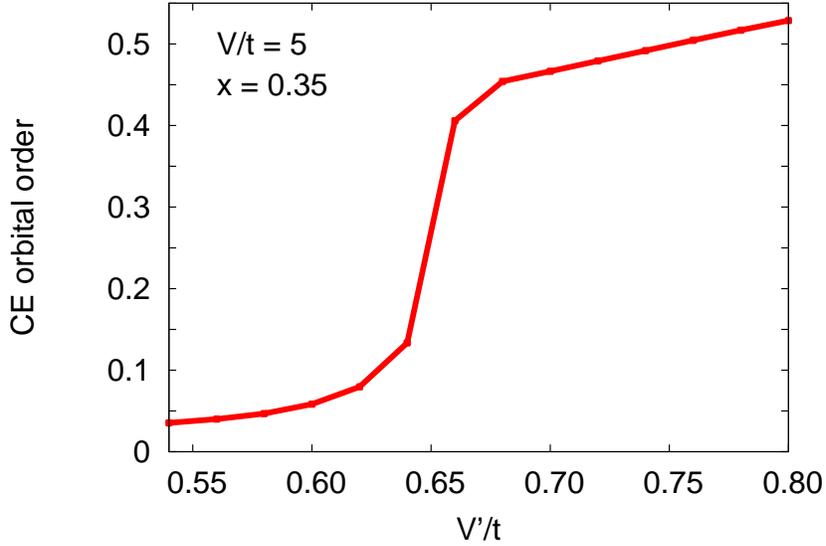,width=110mm}
\vspace{-5mm}
\end{center}
\caption{The sharp sensitivity of the CE-type orbital order to the inter-site charge interaction strength $V'$.}
\label{fig3}
\end{figure}

Fig. \ref{fig3} shows the sharp onset of CE-type orbital order with increasing $V'$, showing its strong sensitivity to the inter-site interaction strength. Here the inter-orbital interaction strength is fixed at $V/t=5$ and hole doping $x=0.35$.

\begin{figure}
\begin{center}
\vspace*{-2mm}
\hspace*{-0mm}
\psfig{figure=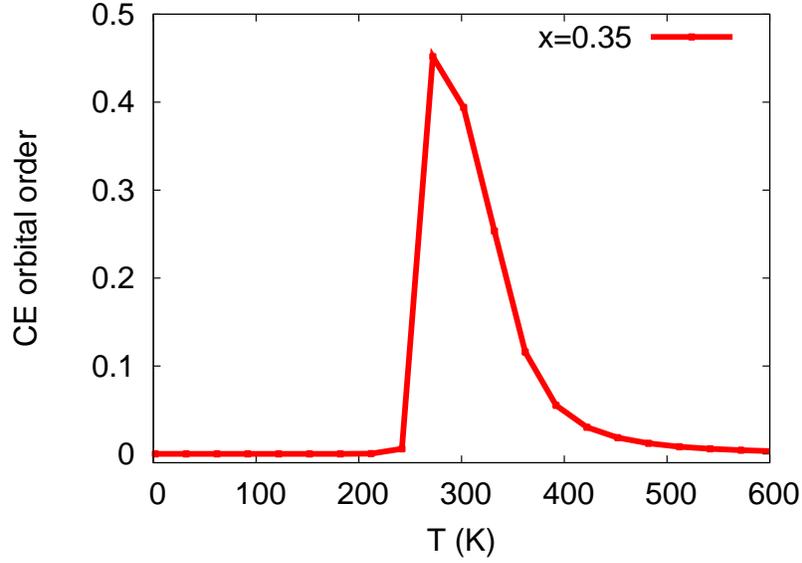,width=110mm}
\vspace{-5mm}
\end{center}
\caption{The sharp onset and melting of the CE-type orbital order close to the Curie temperature in the correlated 
ferromagnet. The shape and temperature scale are quantitatively very similar to the observed behaviour of the $(\pi/2,\pi/2,0)$ intensity in neutron scattering experiments.}
\label{fig4}
\end{figure}

\begin{figure}
\begin{center}
\vspace*{-2mm}
\hspace*{-0mm}
\psfig{figure=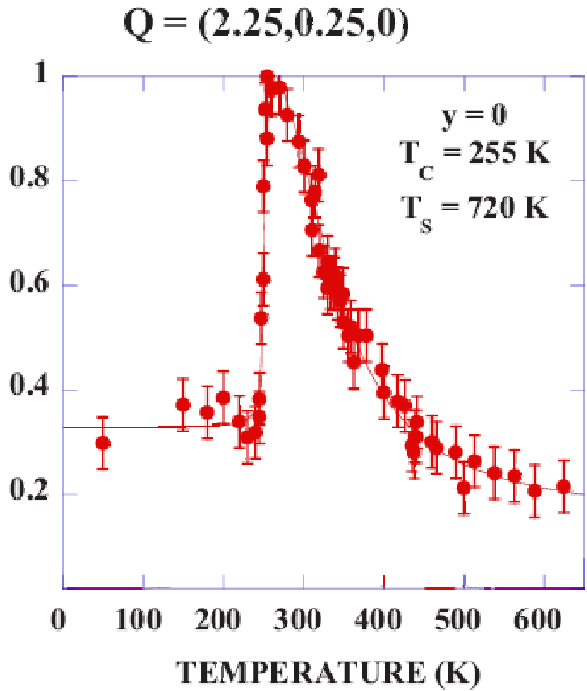,width=110mm}
\vspace{-5mm}
\end{center}
\caption{Behaviour of the $(\pi/2,\pi/2,0)$ intensity observed in neutron scattering experiments on 
$\rm La_{0.7} Ca_{0.3} Mn O_3$ single crystal (from ref. [15]).}
\label{moussa}
\end{figure}

Fig. \ref{fig4} shows the temperature dependence of the local CE order. Due to the extreme sensitivity of CE correlations seen above, a slight enhancement in the ratios $V/t$ and $V'/t$ due to reduction in hopping $t(T)$ with temperature in the correlated ferromagnet results in a sharp onset of CE-type orbital order close to the Curie temperature, which is seen to rapidly melt away with increasing temperature. Both the shape and the temperature scale are in good agreement with observed behaviour of local CE correlations in neutron scattering studies of $\rm La_{0.7} (Ca_{y} Sr_{1-y})_{0.3} Mn O_3$ single crystals,\cite{moussa_2007} as shown in Fig. \ref{moussa} for $y=1$. Here we have used the same hopping energy scale $t(0)=200$meV for manganites as in our recent comparison of spin dynamics results with experiments.\cite{qfklm2}

The CE-type orbital order shown above was obtained using the unrestricted HF scheme of Eq. (12), extended to finite temperature by summing over all states with appropriate Fermi functions. The hopping reduction was included approximately as $t(T)/t(0) = 1 - (1/3)*(1 - \langle S_z \rangle /S )$ in terms of the temperature-dependent magnetization $\langle S_z \rangle$; this yields a reduction from 1 in the ferromagnetic state to 2/3 in the paramagnetic state, which is similar to the hopping reduction by the factor $\langle \cos (\theta_{ij}/2) \rangle$ in the double-exchange model.\cite{millis_1995}
The corresponding thermal enhancements for the ratios $V/t$ and $V'/t$ considered in our self-consistent analysis were from 3.3 to 5.0 and 0.5 to 0.75, respectively. 


Bandwidth (hopping) reduction in a correlated ferromagnet follows from electronic self energy correction due to electron-magnon interaction, which becomes important near $T_c$ due to thermal excitation of local zone-boundary magnons. Due to this band narrowing near the Curie temperature, finite Jahn-Teller distortion and orbital correlations were shown to be self-consistently generated in a two-orbital FKLM.\cite{stier_2007} However, only ferro orbital correlations were considered. 

\begin{figure}
\begin{center}
\vspace*{-2mm}
\hspace*{-0mm}
\psfig{figure=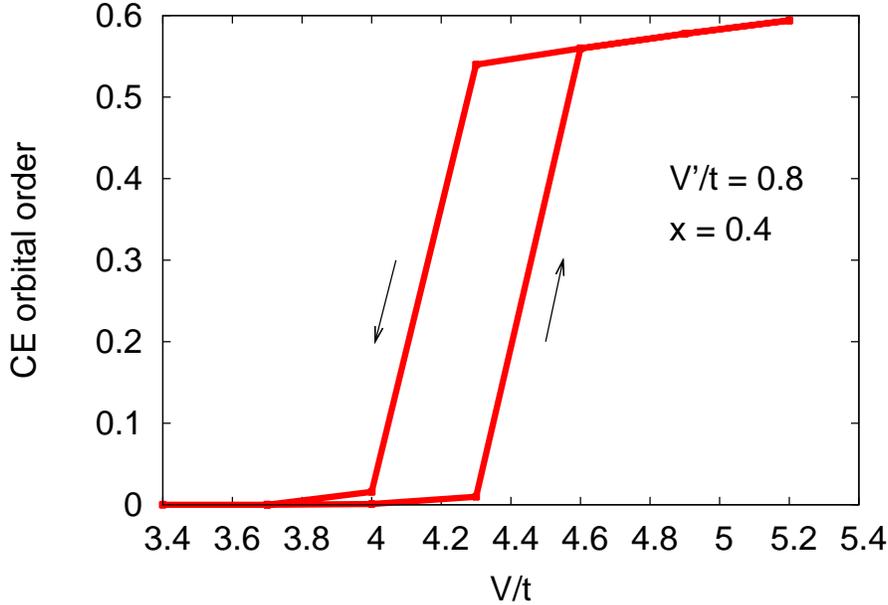,width=120mm}
\vspace{-5mm}
\end{center}
\caption{Onset of CE-type orbital order shows hysteresis behaviour on increasing/decreasing $V/t$ values.}
\label{hyst}
\end{figure}

Fig. \ref{hyst} shows hysteresis in the onset of CE-type orbital order with increasing/decreasing $V/t$. For the same value of $V/t$ (in the range between 4.0 and 4.6), there are two distinct self-consistent states with and without orbital ordering, depending on the history. Bandwidth reduction with temperature will translate this into a hysteresis behaviour with temperature, indicating metastability and coexisiting regions with and without local orbital correlations. This behaviour of CE-type orbital correlations should be important in view of recent observations of spin-glass behaviour, phase separation, and evidence of metastability in manganites near half doping.\cite{chaddah_2008}

\section{CE-type orbital order in zig-zag AF state}

Long-range charge and orbital ordering sets in at half doping in bilayer and pseudo-cubic manganites where equal numbers of nominal $\rm Mn^{3+}$/$\rm Mn^{4+}$ ions form a checkerboard arrangement in a plane with ferromagnetic zigzag chains coupled antiferromagnetically, which is repeated in the perpendicular direction. Evidence for such ordering comes from superstructure reflections at wavevector ($\pi,\pi,0$) and ($\pi/2,\pi/2,0$) in diffraction experiments corresponding to charge and orbital order respectively.

To investigate the onset of local CE-type orbital ordering in such zig-zag AF states in two dimensions, we consider a 4+4 sublattice basis for the spin up and spin down ferromagnetic zig-zag chains. In the HF approximation, the effective single-particle Hamiltonian for spin-$\uparrow$ electrons is extended to: 

\begin{equation}
H^{(0)}  =   \sum_{{\bf k},\mu} \Psi_{{\bf k}\mu} ^\dagger
\begin{pmatrix} 
{\rm H}_{\rm uu} & {\rm H}_{\rm ud} \\
{\rm H}^\dag _{\rm ud} & {\rm H}_{\rm dd} 
\end{pmatrix} 
\Psi_{{\bf k}\mu}
\end{equation}
where the intra-chain term for the spin-$\sigma$ ferromagnetic chain ($\sigma = {\rm up/down}$) is given by:
\begin{equation}
{\rm H}_{\sigma\sigma} ({\bf k})=  
\begin{pmatrix} 
\Delta_{\rm ch} - \mu \Delta_{\rm orb} & \delta_{k_x} & 0 & \delta^{\star}_{k_y} \\
\delta^{\star}_{k_x} & -\Delta_{\rm ch} & \delta_{k_x} & 0 \\
0 & \delta^{\star}_{k_x} & \Delta_{\rm ch} + \mu \Delta_{\rm orb} & \delta_{k_y} \\
\delta_{k_y} & 0 & \delta^{\star}_{k_y} & -\Delta_{\rm ch} 
\end{pmatrix} 
-\sigma J S {\bf 1}
\end{equation}
in the four-sublattice basis introduced earlier, and the inter-chain term:
\begin{equation}
{\rm H}_{\rm ud}({\bf k})=  
\begin{pmatrix} 
0 & \delta_{k_y}  & 0 & \delta_{k_x}^{\star}  \\
\delta^{\star}_{k_y} & 0 & \delta_{k_y} & 0 \\
0 & \delta^{\star}_{k_y} & 0 & \delta_{k_x} \\
\delta_{k_x} & 0 & \delta^{\star}_{k_x} & 0 
\end{pmatrix}
\end{equation}
where $\delta_{k_x}$ etc. are the corresponding components of the NN hopping term given in Eq. (11). The resulting $8 \times 8$ Hermitian matrix was diagonalized to obtain the eigenvalues and eigenvectors, and the charge-orbital ordering was obtained self-consistently as described in the previous section.

\begin{figure}
\begin{center}
\vspace*{-2mm}
\hspace*{-0mm}
\psfig{figure=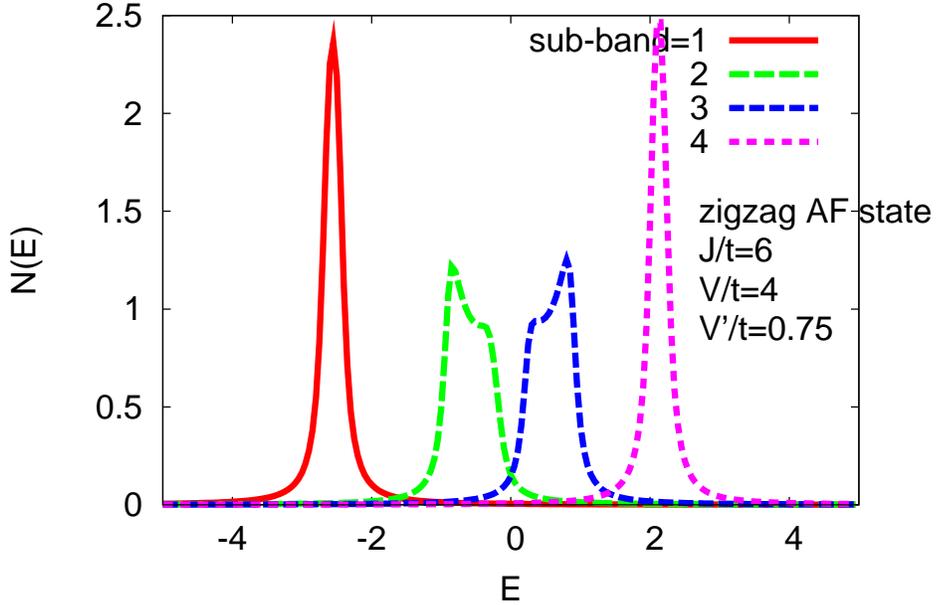,width=120mm}
\vspace{-5mm}
\end{center}
\caption{The spin-$\uparrow$ electronic DOS showing four low-energy sub-bands corresponding to CE-type orbital ordering in the zig-zag AF state. A similar structure shifted up by the exchange splitting $2JS$ corresponding to the spin-down ferromagnetic zig-zag chains is not shown.}
\label{dos_zigzag}
\end{figure}

As the CE-type orbital order involves a four-sublattice structure, Fig. \ref{dos_zigzag} shows the corresponding four sub-bands in the low-energy part of the $e_g$ electronic DOS. There is a similar structure shifted up by the exchange splitting $2JS$ corresponding to the spin-down ferromagnetic chains. Out of the four sub-bands, the lowest one is fully occupied at $x=0.5$ (corresponding to quarter filling in the  spin-$\uparrow$ electron sector).

\begin{figure}
\begin{center}
\vspace*{-2mm}
\hspace*{-0mm}
\psfig{figure=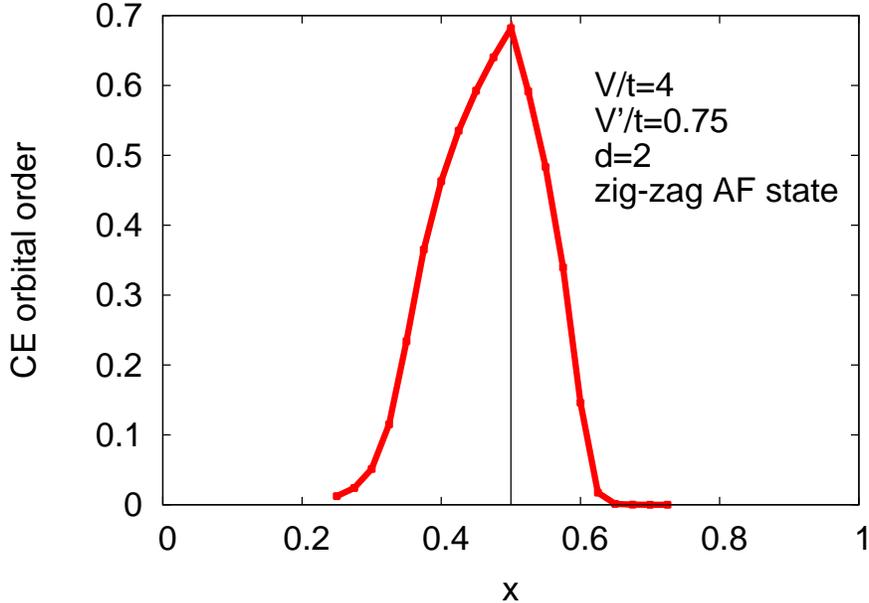,width=120mm}
\vspace{-5mm}
\end{center}
\caption{The CE-type orbital order in a two-dimensionsal zig-zag AF state is stabilized only within a narrow doping range around $x=0.5$.}
\label{zigzag}
\end{figure}

Fig. \ref{zigzag} shows that the CE-type orbital order in the zig-zag AF state is stabilized within a narrow doping range around $x=0.5$. With increasing Hund's coupling $J$ and Coulomb barrier in the AF state, the dimensionality of the mobile $e_g$ electrons decreases effectively from two to one. Due to this reduced dimensionality and delocalization, the CE-type orbital near half doping is found to be more readily stabilized even for smaller interaction strength. Similar doping behaviour of the CE-type orbital order was recently obtained using the Jahn-Teller approach.\cite{schlottmann_2009} 

\section{Conclusions}
A strong sensitivity of local orbital order on doping concentration, interaction strengths, and temperature is revealed in our investigation of onset and melting of local orbital order within a two-orbital model with effective inter-orbital and inter-site Coulomb interactions. 

In the ferromagnetic state, the CE-type orbital order was found to be optimal around $x=0.4$ and to diminish rapidly at $x=0.5$. In the doping regime $0.3 < x < 0.4$, the orbital order was found to rise sharply around $V/t=4$ and $V'/t=0.75$, rendering the system extremely sensitive to small changes in bandwidth and temperature. A sharp thermal onset and melting of this orbital ordering was found, remarkably similar in shape and temperature scale to neutron-scattering observations, when a small temperature-dependent reduction of the hopping term was included within a self-consistent treatment of the local orbital order, with similar values of $t$ and $V$ as used in a recent detailed comparison of calculated spin dynamics for manganites with experiments.\cite{qfklm2}  

The onset of CE-type orbital order was found to exhibit hysteresis behaviour on increasing/decreasing $V/t$ (temperature), with two distinct self-consistent states with and without orbital ordering. Indicating metastability and coexisiting regions with and without local orbital correlations, this hysteresis behaviour should be important in view of recent observations of spin-glass behaviour, phase separation, and evidence of metastability in manganites near half doping.\cite{chaddah_2008}

In a two-dimensional CE-type AF state, with alternating zig-zag ferromagnetic chains, the CE-type orbital order was found to be stabilized only within a very narrow doping range around $x=0.5$. Due to reduced dimensionality and delocalization of mobile $e_g$ electrons in the zig-zag AF state, the CE-type orbital order near half doping is stabilized for even smaller interaction strengths. This strong sensitivity and its rapid melting away on either electron or hole doping implies that CE-type orbital order would be highly susceptible in a strong magnetic field as well, which is of interest in context of magnetic field induced melting of CE-type orbital order in half-doped manganites.\cite{mukherji_2009}

While only local orbital correlations (local moments in pseudo-spin space) were considered here, which are associated with short time scale correlations corresponding to the high-energy scale $V$, including low-energy orbital fluctuations would allow for investigation of long-range orbital ordering features (orbital correlation length, orbital disordering temperature) and reduction of orbital order due to quantum and thermal excitation of orbitons. Also, renormalization of the CE-state electron spectral properties due to electron-orbiton coupling self energy resulting from multiple orbiton emission/absorption processes would provide insight into correlated lattice polarons.


\begin{thebibliography}{08}

\bibitem{tokura_2003} 
Y. Tokura, Physics Today, 50, (2003). 

\bibitem{khaliullin_2005}
G. Khaliullin, Prog. Theor. Phys. Suppl. {\bf 160}, 155 (2005). 

\bibitem{saitoh_2001}
E. Saitoh, S. Okamoto, K. T. Takahashi, K. Tobe, K. Yamamoto, T. Kimura, S. Ishihara, S. Maekawa and Y. Tokura,
Nature {\bf 410}, 180 (2001).

\bibitem{ogasawara_2001} 
T. Ogasawara, T. Kimura, T. Ishikawa, M. Kuwata-Gonokami, and Y. Tokura, Phys. Rev. B {\bf 63}, 113105 (2001).

\bibitem{khalifah_2002}
P. Khalifah, R. Osborn, Q. Huang, H. W. Zandbergen, R. Jin, Y. Liu, D. Mandrus, R. J. Cava, 
Science, {\bf 297}, 2237 (2002). 

\bibitem{ulrich_2008}
C. Ulrich, G. Ghiringhelli, A. Piazzalunga, L. Braicovich, N. B. Brookes, H. Roth, T. Lorenz, and B. Keimer,
Phys. Rev. B {\bf 77}, 113102 (2008). 

\bibitem{lee_2009}
W.-C. Lee and C. Wu, Phys. Rev. Lett. {\bf 103}, 176101 (2009).

\bibitem{wollan_1955}
E. O. Wollan and W. C. Koehler,
Phys. Rev. {\bf 100}, 545 (1955).

\bibitem{goodenough_1955}
J. B. Goodenough,
Phys. Rev. {\bf 100}, 564 (1955).

\bibitem{argyriou_2002}
D. N. Argyriou, J.W. Lynn, R. Osborn, B. Campbell, J. F. Mitchell, U. Ruett, H. N. Bordallo, A. Wildes and C. D. Ling,
Phys. Rev. Lett. {\bf 89}, 3954 (2002)

\bibitem{campbell_2003}
B. J. Campbell, S. K. Sinha, R. Osborn, S. Rosenkranz, J. F. Mitchell, D. N. Argyriou, L. Vasiliu-Doloc,
O. H. Seeck and J. W. Lynn,
Phys. Rev. B {\bf 67}, 020409 (2003)

\bibitem{adams_2000}
C. P. Adams, J.W. Lynn, Y. M. Mukovskii, A. A. Arsenov and D. A. Shulyatev,
Phys. Rev. Lett. {\bf 85}, 3954 (2000).

\bibitem{lynn_2007}
J. W. Lynn, D. N. Argyriou, Y. Ren, Y. Chen, Y. M. Mukovskii and D. A. Shulyatev, Phys. Rev. B {\bf 76}, 014437 (2007).

\bibitem{saitoh_2002}
E. Saitoh, Y. Tomioka, T. Kimura, and Y. Tokura, 
J. Magn. Magn. Mater. {\bf 239}, 170 (2002). 

\bibitem{moussa_2007}
F. Moussa, M. Hennion, P. Kober-Lehouelleur, D. Reznik, S. Petit, H. Moudden, A. Ivanov, Ya. M. Mukovskii, R. Privezentsev and F. Albenque-Rullier,
Phys. Rev. B {\bf 76}, 064403 (2007).

\bibitem{tomioka_2000}
Y. Tomioka, A. Asamitsu, and Y. Tokura, Phys. Rev. B {\bf 63}, 024421 (2000). 

\bibitem{sporb}
D. K. Singh, B. Kamble, and A. Singh, 
Phys. Rev. B {\bf 81}, 064430 (2010).

\bibitem{qfklm2}
D. K. Singh, B. Kamble, and A. Singh, 
J. Phys.: Condens. Matter {\bf 22}, 396001 (2010).

\bibitem{hwang_98}
H. Y. Hwang, P. Dai, S-W. Cheong, G. Aeppli, D. A. Tennant, and H. A. Mook, Phys. Rev. Lett. {\bf 80}, 1316 (1998).

\bibitem{dai_2000}
P. Dai, H. Y. Hwang, J. Zhang, J. A. Fernandez-Baca, S.-W. Cheong, C. Kloc, Y. Tomioka, and  Y. Tokura, 
Phys. Rev. B {\bf 61}, 9553 (2000).

\bibitem{tapan_2002}
T. Chatterji, L. P. Regnault, and W. Schmidt, Phys. Rev. B {\bf 66}, 214408 (2002).

\bibitem{ye_2006}
F. Ye, Pengcheng Dai, J. A. Fernandez-Baca, Hao Sha, J. W. Lynn, H. Kawano-Furukawa, Y. Tomioka, Y. Tokura, and Jiandi Zhang, Phys. Rev. Lett. {\bf 96}, 047204 (2006).

\bibitem{ye_2007}
F. Ye, P. Dai, J. A. Fernandez-Baca, D. T. Adroja, T. G. Perring, Y. Tomioka, and Y. Tokura, 
Phys. Rev. B {\bf 75}, 144408 (2007) .

\bibitem{zhang_2007}
J. Zhang, F. Ye, H. Sha, P. Dai, J. A. Fernandez-Baca, and E. W. Plummer,
J. Phys.: Condens. Matter {\bf 19}, 315204 (2007).

\bibitem{millis_1995}
A. J. Millis, P. B. Littlewood and B. I. Shraiman,
Phys. Rev. Lett. {\bf 74}, 5144 (1995).

\bibitem{benedetti_1999}
P. Benedetti and R. Zeyher,
Phys. Rev. B {\bf 59}, 9923 (1999).

\bibitem{okamoto_2002}
S. Okamoto, S. Ishihara, and S. Maekawa,
Phys. Rev. B {\bf 65}, 144403 (2002).

\bibitem{hotta_2000}
T. Hotta, A. L. Malvezzi, and E. Dagotto,
Phys. Rev. B {\bf 62}, 9432 (2000).

\bibitem{stier_2007}
M. Stier and W. Nolting, 
Phys. Rev. B {\bf 75}, 144409 (2007).

\bibitem{maitra_2003}
T. Maitra and A. Taraphder, Phys. Rev. B {68}, 174416 (2003).  

\bibitem{pissas_2005}
M. Pissas, I. Margiolaki, G. Papavassiliou, D. Stamopoulos, and D. Argyriou,
Phys. Rev. B {\bf 72}, 064425 (2005).

\bibitem{as_af_ordering}
A. Singh and Z. Te\v{s}anovi\'{c}, Phys. Rev. B {\bf 41}, 614 (1990);
A. Singh, Phys. Rev. B {\bf 43}, 3617 (1991);
A. Singh, cond-mat/9802047 (1998).

\bibitem{as_doped_af}
A. Singh and H. Ghosh, Phys. Rev. B {\bf 65}, 134414 (2002).


\bibitem{hotta_2001}
T. Hotta, A. Feiguin and E. Dagotto, 
Phys. Rev. Lett. {\bf 86}, 4922 (2001).

\bibitem{cepas_2006}
O. Cepas, H. R. Krishnamurthy, and T. V. Ramakrishnan, Phys. Rev. B {\bf 73}, 035218 (2006).

\bibitem{schlottmann_2009}
P. Schlottmann, Phys. Rev. B {\bf 80}, 104428 (2009).

\bibitem{sboychakov_2010}
A. O. Sboychakov, K. I. Kugel, A. L. Rakhmanov, and D. I. Khomskii, arXiv:1007.4814 (2010).

\bibitem{shen_2010}
C. \c{S}en, G. Alvarez, and E. Dagotto, Phys. Rev. Lett. {\bf 105}, 097203 (2010).

\bibitem{rosciszewski_2010}
K. Ro\'{s}ciszewski and A. M. Ole\'{s}, J. Phys.: Condens. Matter {\bf 22}, 425601 (2010).


\bibitem{chaddah_2008}
P. Chaddah, K. Kumar and A. Banerjee,
Phys. Rev. B {\bf 77}, 100402 (2008).

\bibitem{mukherji_2009}
A. Mukherji and P. Majumdar, arXiv:0811.3563 (2009), and references therein.

\end{thebibliography}
\end{document}